\ifx\fiverm\undefined
        \newfont\fiverm{cmr5} 
\fi
\input prepictex
\input pictex
\input postpictex
\catcode`\@=11
\@input{picmore.tex}
\documentstyle[11pt]{article}

\def\labelmark{}

\def\void{}
\newenvironment{formula}[1]{\def\labelname{#1}
\ifx\void\labelname\def\junk{\begin{displaymath}}
\else\def\junk{\begin{equation}\label{\labelname}}\fi\junk}%
{\ifx\void\labelname\def\junk{\end{displaymath}}
\else\def\junk{\end{equation}}\fi\junk\labelmark\def\labelname{}}
{\ifx\void\labelname\def\junk{\end{array}\end{displaymath}}
\else\def\junk{\end{array}\right.\end{equation}}
\fi\junk\labelmark\def\labelname{}\def\junk{}
}
\newenvironment{formulae}[1]{\def\labelname{#1}
\ifx\void\labelname\def\junk{\begin{displaymath}}
\else\def\junk{\begin{eqnarray}\label{\labelname}}\fi\junk}%
{\ifx\void\labelname\def\junk{\end{displaymath}}
\else\def\junk{\end{eqnarray}}\fi\junk\labelmark\def\labelname{}}

\newcommand{\beq}{\begin{formula}}
\newcommand{\eeq}{\end{formula}}
\newcommand{\beqa}{\begin{formulae}}
\newcommand{\eeqa}{\end{formulae}}
\newcommand{\nn}{\nonumber}
\newcommand{\eq}[1]{(\ref{#1})}

\newcommand{\define}{\stackrel{\mbox{def.}}{=}}
\newfont{\sgcal}{eufm9}
\newfont{\smcal}{eusm9}
\newfont{\smbf}{msbm9}
\newfont{\gcal}{eufm10}
\newfont{\mcal}{eusm10}
\newfont{\mbf}{msbm10}
\newfont{\Gcal}{eufm10 scaled\magstep1}
\newfont{\Mcal}{eusm10 scaled\magstep1}
\newfont{\Mbf}{msbm10 scaled\magstep1}
\newcommand{\ra}{\rightarrow}
\newcommand{\lra}{\leftrightarrow}
\newcommand{\der}{\partial}
\newcommand{\eps}{\varepsilon}
\newcommand{\dand}{\hskip-40pt &&}
\newcommand{\Fra}{\mbox{\Gcal a}}
\newcommand{\Frb}{\mbox{\Gcal b}}
\newcommand{\Z}{\mbox{\Mbf Z}}
\newcommand{\z}{\mbox{\smbf Z}}
\newcommand{\NP}[1]{ {\it Nucl.~Phys.} {\bf #1}}
 
\newcommand{\PL}[1]{ {\it Phys.~Lett.} {\bf #1}}

\newcommand{\PR}[1]{ {\it Phys.~Rev.} {\bf #1}}

\newcommand{\PTP}[1]{ {\it Prog.~Theor.~Phys.} {\bf #1}}

\newcommand{\MPL}[1]{ {\it Mod.~Phys.~Lett.} {\bf #1}}
\newcommand{\IJMP}[1]{ {\it Int.~J.~Mod.~Phys.} {\bf #1}}
\newcommand{\JP}[1]{ {\it J.~Phys.} {\bf #1}:\  Math.~Gen.~}
\newcommand{\ZP}[1]{ {\it Z.~Phys.} {\bf #1}}

\newcommand{\AP}[1]{ {\it Ann.~Phys. (N.Y.)} {\bf #1}}
\newcommand{\JMP}[1]{ {\it J. Math.~Phys.} {\bf #1}}
\newcommand{\WS}[1]{ (World Scientific Publishing, #1)}

\begin{document}
\begin{titlepage}
\setcounter{page}{1}
\renewcommand{\thefootnote}{\fnsymbol{footnote}}


\begin{flushright}
\end{flushright}

\vspace{3mm}
\begin{center}
{\Large Deformed fields and Moyal construction of deformed super Virasoro algebra} 
\vspace{9mm}

{\large Ryuji Kemmoku$^{1)}$
\footnote{E-mail: kemmoku@sci.kitasato-u.ac.jp }}
\,\sc{and}\,
{\large Haru-Tada Sato$^{2)}$   
\footnote{E-mail: haru@taegeug.skku.ac.kr }}\\
\vspace{3mm}
\vspace{3mm}
{\em $^{1)}$ Nonlinear Physics Laboratory, Department of Physics\\
School of Science, Kitasato University\\
Sagamihara 228-8555, Japan}\\
\vspace{5mm}
{\em $^{2)}$ Department of Physics, Sungkyunkwan 
University, Suwon 440-746, Korea}
\end{center}

\vspace{5mm}

\begin{abstract}
Studied is the deformation of super Virasoro algebra proposed by Belov and Chaltikian. 
Starting from abstract realizations in terms of the FFZ type generators, various connections 
of them to other realizations are shown, especially to deformed field 
representations, whose bosonic part generator is recently reported as a deformed 
string theory on a noncommutative world-sheet. The deformed Virasoro generators can 
also be expressed in terms of ordinary free fields in a highly nontrivial way.  
\end{abstract}

\vspace{5mm}
\vfill
\begin{flushleft}
\end{flushleft}

\end{titlepage}
\newpage
\setcounter{footnote}{0}
\renewcommand{\thefootnote}{\arabic{footnote}}
\renewcommand{\theequation}{\thesection.\arabic{equation}}

\section{Introduction}\label{sec1}
\setcounter{section}{1}
\setcounter{equation}{0}
\indent


The deformed Virasoro algebras proposed in an early 
stage~\cite{saito}-~\cite{N2} have recently become more suitable to be 
examined in physical connections. These algebras were studied originally 
from the purely mathematical motivation to seek a $q$-analogue of the 
Virasoro algebra in the context of quantum groups. However, as far as 
these algebras are concerned, it has become more appropriate to discuss 
in some physical connections than in the quantum group context; for example, 
in string theory~\cite{QST}, a solvable lattice model~\cite{QKZ}, the 
lowest Landau levels~\cite{me,Hall}, and soliton systems~\cite{KP}. 

Among them a bosonic string on a noncommutative geometry has recently 
been constructed~\cite{QST}, and its string action possesses the deformed 
Virasoro algebra as a symmetry, like the Virasoro algebra for the ordinary 
bosonic string (on commutative world-sheets). The field equation obeys a 
discrete time evolution, and the oscillators satisfy the usual $q$-deformed 
Heisenberg algebra, 
\beq{qboson}
[a_n,a_m]={q^n-q^{-n}\over q-q^{-1}}\,\delta_{n+m,0}  \ . 
\eeq
Noncommutative geometry is suggested to appear in a particular lowest 
energy limit of string theory in a constant background $B$-field 
configuration~\cite{CH,SW}. This situation is similar to the appearance of 
noncommutative discrete translations (magnetic translations~\cite{mag}) 
for electron systems confined on a two-dimensional surface in a constant 
magnetic field. In these lowest Landau level systems, the same deformed 
Virasoro algebra is realized~\cite{me} as a particular combination of the 
Fairlie-Fletcher-Zachos (FFZ) algebra~\cite{FFZ} (or algebra of magnetic 
translations). The discrete nature of a magnetic lattice is also a 
resemblance to the discrete time prescription of the above noncommutative 
string theory. In fact, the FFZ type realization implies an intimate 
relation to the Moyal type deformation~\cite{moyal}, which has recently 
been paid much attention in the context of noncommutative string/field 
theories~\cite{NC}. 

In order to find the corresponding deformed symmetry of a superstring on 
a noncommutative geometry in view of the above connection, it is worth to 
first realize a supersymmetric extension of the deformed Virasoro algebra 
in terms of the FFZ generators. In this paper, we focus our attention on 
the superalgebra proposed by Belov and Chaltikian (BC)~\cite{BC}. This 
superalgebra is in fact a supersymmetric extension of the symmetry in the 
deformed bosonic string~\cite{QST}. We shall follow the same methods as 
developed in a different type of deformed super Virasoro 
algebra~\cite{me,OPE1,OPE2}, where bosonic and fermionic parts 
participate in an asymmetric way. In the present case, they are 
symmetric, and this is certainly an advantageous point. However, magnetic 
translations are nothing but differential operators ($q$-difference 
operators), and the deformed super Virasoro algebra obtained in this way 
becomes a centerless algebra. {}~From the viewpoint of applications to 
field theories, we hence have to make a connection from the centerless 
realizations to a field realization. Hence, starting from the most general 
formulae, we shall derive various realizations and their related 
formulae extensively in terms of differential operators, matrices, 
ordinary/deformed fields (boson, fermion, and ghosts). 

One of nontrivial issues of the paper is the relation between ordinary and 
deformed fields. In this paper we introduce the deformed boson field which is 
comprised of the $q$-boson oscillators satisfying the relation~\eq{qboson}. 
(We also introduce a similar deformed fermion.) Taking account of normalization 
changes in \eq{qboson}, the ($q$-boson) oscillators look equivalent to the usual 
boson oscillators. However, as suggested in ~\cite{QST}, the parameter $q$ 
possesses the clear meaning of a time discretization on a world-sheet, emerged 
from a noncommutative geometry. We shall also support the nontriviality of 
the relation \eq{qboson} from a different viewpoint. Although the commutation 
relation itself may be trivial in the sense of normalization, it is no longer 
true at the level of a field object. We suggest the ordinary field realizations 
which cannot be connected to the deformed fields by the normalization changes. 
Even if connected to the deformed fields in such a way, the deformed fields 
become nonlocal objects which require fractional differential calculus. 
Although the present formulation is not directly related to noncommutative 
geometry, the symmetry generator form is exactly the same as in the deformed 
bosonic string~\cite{QST} (see also \cite{QKZ}, \cite{OPE2,CP2}). There should 
be clear connections among them, and hence it is very useful to study the 
corresponding superalgebra. 

The deformed Virasoro algebra may also play an important role in 
integrable systems. To reveal the nature of integrability, it is 
of course necessary to understand what type of deformation maintains the 
integrability in each system. Otherwise pathological deformation tends 
to destroy integrability leading to a chaotic behavior. 
{}~From this view point, magnetic translations are suitable mathematical 
means of describing an integrable discretization, since it is a 
well-behaving difference operator on a lattice. 
In fact, the deformed Virasoro algebra (noncommutative magnetic translation) 
on a magnetic lattice becomes the Virasoro algebra (commutative continuous 
translation) as a magnetic field vanishes. (This is exactly the $q\ra1$ 
limit~\cite{LLL} in the terminology of quantum groups.) 
We hence expect that a system possessing the deformed Virasoro 
symmetry will be related to its integrability in somehow algebraic way. 
Similarly, the investigation on this symmetry might provide novel 
suggestion in various related areas as well as in string theory. 

This paper is organized as follows. In Section~\ref{sec2}, we put necessary 
formulae and brief comments on the BC superalgebra. Section~\ref{sec3} is an 
entirely new part, where we discuss the deformed field realizations of the BC 
superalgebra. We present bilinear integral forms, adjoint commutator 
representations and their matrix forms.  In Section~\ref{sec4}, we show two FFZ 
realizations of the (centerless) BC superalgebra. 
When describing them by differential operators and the Pauli matrices, 
both realizations are organized into the similar (but slightly different) matrix 
forms presented in Section~\ref{sec3}. In Section~\ref{sec5}, we improve this 
different matrix structure by introducing another set of the FFZ generators as 
well as a particular noncommutative generalization of the Pauli matrices. We show 
four classes of the realizations of this type. Each class is an infinite set 
represented by one parameter $\Delta$. Among them we only discuss two specific 
cases, which exactly reduce to the same matrix forms as presented in 
Section~\ref{sec3}. These are discussed case by case. It is also shown that the 
familiar differential realizations in superspace are obtained in the limit 
of $q\ra1$. In Section~\ref{sec6}, we discuss the realizations by ordinary 
free fields. Their relations to the deformed field realizations are highly 
nontrivial. Section~\ref{sec7} concerns (super) ghost field realizations. 
We obtain the same copies of the BC algebra with new central extensions. 
Section~\ref{sec8} contains conclusions and discussions.

\section{The Belov-Chaltikian (BC) superalgebra}\label{sec2}
\setcounter{section}{2}
\setcounter{equation}{0}
\indent 

The superalgebra proposed by Belov and Chaltikian~\cite{BC} is an 
algebra with two sets of indices~\footnote{Their supercurrent generators 
are related to our $G_r^{(k)}$ by the relation 
$ {\cal F}^{k,\pm}_r={1\over2}(G_r^{(-k)}\pm G_r^{(k)} ) $.}
, and organized in the following way (see \cite{OPE2} for more details):
\beq{BCalg}
[\, L_n^{(k)}\,,L_m^{(l)}\,]={1\over2}\sum_{\eps,\eta=\pm1}
\biggl[{n(\eps l+1)-m(\eta k+1)\over2}\biggr]_- 
L_{n+m}^{(\eps k+\eta l +\eps\eta)} +C\delta_{n+m,0}\ .
\eeq
\beq{LG}
[\,L_n^{(k)}\,,G_m^{(l)}\,] ={1\over2}\sum_{\eps,\eta=\pm1}\eps
\left({2\over q-q^{-1}}\right)^{1-\eps\eta\over2}
\Bigl[ {n(l+\eta)-m(\eps k+\eta)\over2}\Bigr]_{-\eps\eta}
G^{(l+\eps k+\eta)}_{n+m} \ ,
\eeq
\beq{GG0}
\{\,G_r^{(k)}\,,G_s^{(l)}\,\} = 
\sum_{\eps,\eta=\pm1}
\left({q-q^{-1}\over 2}\right)^{1-\eps\eta\over2}
\Bigl[ {r(l-\eps)+ s(k+\eps)\over2}\Bigr]_{\eps\eta}
L^{(\eta k-\eta l+\eps\eta)}_{r+s}\, +C_G\delta_{r+s,0}\ ,
\eeq
where we define 
\beq{.0}
[x]_+=(q^x+q^{-x})/2\ ,\qquad\,  
[x]_-=(q^x-q^{-x})/(q-q^{-1}) \ ,
\eeq
or in a symbolic way
\beq{xpm}
[x]_\eps =\left({2\over q-q^{-1}}\right)^{1-\eps\over2}
{q^x+ \eps q^{-x}\over2}\ . \qquad\quad (\eps=\pm1)
\eeq
The indices ($n$ and $k$) on $L_n^{(k)}$ run over all integers. 
The lower index on $G_r^{(k)}$ runs half-integers 
for the Neveu-Schwarz type algebra, and integers for the Ramond type. 
The upper index on $G_r^{(k)}$ is an integer. 

The central extensions $C$ and $C_G$ are given as follows~\cite{OPE2}:
\beq{C}
         C = C_B + C_H \ ,
\eeq
with $C_B$ for a scalar field,
\beq{CB}
C_B= {1\over2}\sum_{j=1}^n [k({n\over2}-j)]_+ [l({n\over2}-j)]_+
[n-j]_- [j]_- \ ,
\eeq
and $C_H$ for a Neveu-Schwarz (NS) fermion,
\beq{CH}
C_H= {1\over2}\sum_{j=1}^n [k({n+1\over2}-j)]_- [l({n+1\over2}-j)]_-
[n-j+{1\over2}]_+ [j-{1\over2}]_+ \ .
\eeq
The $C_H$ for a Ramond fermion is given by interchanging $[x]_+$ and 
$[x]_-$ in the expression $C_B$. The other one $C_G$ is  
\beq{CG}
C_G = \sum_{r\leq j<r+s}q^{(s/2+r-j)l+(r/2-j)k}[j-r]_-[j]_+ \ ,
\eeq
where $j\in\Z+1/2$ for the NS case, and $j\in\Z$ for the R case.

In the following, we consider the scalar and fermion parts separately; 
i.e., splitting
\beq{.1}
L_n^{(k)} = H_n^{(k)} + B_n^{(k)} \ ,
\eeq
with
\beq{.2}
[\,H_n^{(k)}, B_m^{(l)}\,]=0 \ .
\eeq
These two parts, $H_n^{(k)}$ and $B_n^{(k)}$, satisfy the 
algebra \eq{BCalg} with the central extensions $C_B$ and $C_H$ 
respectively, and the following set of decomposed relations is 
consistent with \eq{LG} and \eq{GG0}:
\beq{HG}
[\,H_n^{(k)}, G_r^{(l)}\,]={1\over2(q-q^{-1})}\sum_{\eps,\eta=\pm1}
\eps\, q^{n(l+\eta)-r(\eps k+\eta)\over2}G_{n+r}^{(\eps k+l+\eta)} \ ,
\eeq
\beq{BG}
[\,B_n^{(k)}, G_r^{(l)}\,]={-1\over2(q-q^{-1})}\sum_{\eps,\eta=\pm1}
\eta\, q^{-n(l+\eta)+r(\eps k+\eta)\over2}G_{n+r}^{(\eps k+l+\eta)} \ ,
\eeq
\beq{GG}
\{\,G_r^{(k)}, G_s^{(l)}\,\}=\sum_{\eps=\pm1}
\left( q^{r(l-\eps)+s(k+\eps) \over2} B_{r+s}^{(k-l+\eps)}
+\eps\, q^{-r(l-\eps)-s(k+\eps) \over2} H_{r+s}^{(k-l+\eps)}\right) 
+ C_G \delta_{r+s,0}\ .
\eeq
This is the explicit set of the BC superalgebra that we discuss in this 
paper. There are a few remarks: (i) for the consistency between 
Eqs.\eq{HG}--\eq{GG} and Eqs.\eq{LG} and \eq{GG0}, we need the properties 
\beq{BH-}
B_n^{(-k)}=B_n^{(k)}\ ,\qquad\quad H_n^{(-k)}=-H_n^{(k)}\ .
\eeq
(ii) The possible $q\ra1$ limits of \eq{BCalg} are the following 
two ways: $L_n^{(k)}\ra L_n$ and $L_n^{(k)}\ra k L_n$. Assuming 
the $q\ra1$ limits to be
\beq{lim}
B_n^{(k)}\ra L_n^B\ ,\qquad\quad  H_n^{(k)} \ra k L_n^F\ ,
\eeq
where $L_n^B$ and $L_n^F$ are the usual Virasoro generators with $c=1$ 
and $c=1/2$, the superalgebra \eq{BCalg} with Eqs. \eq{HG}--\eq{GG} 
reproduces the correct super Virasoro algebra. 
Thus, it is very natural to have {\it two different FFZ realizations} 
of the algebra \eq{BCalg}, which satisfies the above limit property, 
as found in \cite{me}. 
(iii) When realizing $H_n^{(k)}$ in terms of fermionic field and related 
differential operators, the $k=0$ modes should be treated as in the limit 
of $H_n^{(k)}/[k]_-$. This seems to be natural from the above limit 
behavior of $H_n^{(k)}$. In contrast, as will be seen later, our FFZ 
realizations do not need this special treatment for the $k=0$ modes. 

\section{The free field realizations}\label{sec3}
\setcounter{section}{3}
\setcounter{equation}{0}
\indent

In this section, using the field realizations, we derive various 
formulae for the BC superalgebra. Let us introduce the following 
deformed free fields ($r\in\Z+{1\over2}$ for NS and $r\in\Z$ for R): 
the fermionic field 
\beq{Psi}
\Psi(z)=\sum_r \Frb_r\, z^{-r-{1\over 2}}\ ,
\qquad\{\Frb_r,\Frb_s\}=[r]_+\,\delta_{r+s,0}\ ,
\eeq
and the bosonic field
\beq{Phi}
\Phi(z)=\sum_{n\in\z} \Fra_n\, z^{-n-1}\ ,
\qquad  [\Fra_n,\Fra_m]=[n]_-\,\delta_{n+m,0}\ .
\eeq
One can also express the bosonic field as 
\beq{*}
\Phi(z)={1\over z}[z\der_z]_-\, \phi(z)\ ,
\eeq
if one introduces the analogue of a nondeformed massless scalar field,
\beq{*0}
\phi(z) = i\phi_0+\Fra_0{q-q^{-1}\over2\ln q}\ln z
-\sum_{n\not=0}{\Fra_n\over[n]_-}z^{-n}\ , \qquad 
[\phi_0,\Fra_n]=i\delta_{n,0} \ .
\eeq
The generators satisfying all the relations presented in 
Section~\ref{sec2} are given by
\beq{H}
H_n^{(k)}=\oint_0{dz\over2\pi i}z^{n+1}H^{(k)}(z)
={1\over 2}\sum_r\Bigl[k({n\over 2}-r)\Bigr]_-\,:\Frb_r\Frb_{n-r}:\ ,
\eeq
\beq{B}
B_n^{(k)}=\oint_0{dz\over2\pi i}z^{n+1}B^{(k)}(z)
={1\over 2}\sum_{j\in\z}\Bigl[k({n\over 2}-j)\Bigr]_+\,:\Fra_j\Fra_{n-j}:\ ,
\eeq
\beq{G}
{G}_r^{(k)}=\oint_0{dz\over2\pi i}z^{r+1/2}G^{(k)}(z)
=\sum_{j\in\z} q^{-k({r\over 2}-j)}\,\Fra_j\Frb_{r-j}\ ,
\eeq
where the current fields are defined as 
\beq{Hz}
H^{(k)}(z) ={1\over z(q-q^{-1})}:\Psi(q^{k/2}z)\Psi(q^{-k/2}z):\ ,
\eeq
\beq{Bz}
B^{(k)}(z) = {1\over 2}:\Phi(q^{k/2}z)\Phi(q^{-k/2}z):\ ,
\eeq
\beq{Gz}
G^{(k)}(z) = q^{-{k\over 4}}\Psi(q^{k/2}z)\Phi(q^{-k/2}z)\ .
\eeq
We calculate the commutation relations between the generators and 
the fields:
\beq{Hpsi1}
[H_n^{(k)},\,\Psi(z)\,] = z^{n}
\left[k\Bigl(z\der_{z} +{n \over 2}+{1\over2}\Bigr) \right]_{-}
         \Bigl[ z\der_{z}+n+{1\over2} \Bigr]_{+}\,\Psi(z)\ ,
\eeq
\beq{Bphi1}
[B_n^{(k)},\,\Phi(z)\,] = z^{n}
\left[k\Bigl(z\der_{z} +{n \over 2}+ 1 \Bigr) \right]_{+}
         \Bigl[ z\der_{z}+n+ 1 \Bigr]_{-}\,\Phi(z)\ ,
\eeq
\beq{Gpsi1}
\{G_r^{(k)},\,\Psi(z)\,\}= z^{r+{1\over2}}
q^{-k(z\der_{z}+{r\over 2}+ 1)}
      \Bigl[ z\der_{z}+r + 1 \Bigr]_+\,\Phi(z)\ ,
\eeq
\beq{Gphi1}
[G_r^{(k)},\,\Phi(z)\,]= z^{r-{1\over2}}
q^{k(z\der_{z}+{r\over 2}+{1\over2})}
\Bigl[ z\der_{z}+r+{1\over2}\Bigr]_- \,\Psi(z)\ .
\eeq
We also find another set of commutator representations 
for an arbitrary parameter $\Delta$;
\beq{Hpsi2}
[H_n^{(k)},{1\over {z^{\Delta-1/2}}}\,\Psi(z)\,] = z^{n}
\left[k\Bigl(z\der_{z} +{n \over 2}+\Delta\Bigr)  \right]_{-}
         \Bigl[ z\der_{z}+n+\Delta \Bigr]_{+}\,
{1\over {z^{\Delta-1/2}}}\,\Psi(z)\ ,
\eeq
\beq{Bphi2}
[B_n^{(k)},{1\over {z^{\Delta -1}}}\,\Phi(z)\,]= z^{n}
\left[k\Bigl(z\der_{z} +{n \over 2}+\Delta \Bigr) \right]_{+}
         \Bigl[ z\der_{z}+n+\Delta \Bigr]_{-}\,
{1\over {z^{\Delta-1}}}\,\Phi(z)\ ,
\eeq
\beq{Gpsi2}
\{G_r^{(k)},{1\over {z^{\Delta-1/2}}}\,\Psi(z)\,\}=
z^r q^{-k(z\der_{z}+{r\over 2}+\Delta)}
   \Bigl[ z\der_{z}+r+\Delta \Bigr]_+\,{1\over {z^{\Delta-1}}}\,\Phi(z)\ ,
\eeq
\beq{Gphi2}
[G_r^{(k)},{1\over {z^{\Delta-1}}}\,\Phi(z)\,]=
z^r q^{k(z\der_{z}+{r\over 2}+\Delta)}\Bigl[ z\der_{z}+r+\Delta\Bigr]_-
\, {1\over {z^{\Delta-1/2}}}\,\Psi(z)\ ,
\eeq
where we notice that the differential operators acting on 
$z^{{1\over2}-\Delta}\Psi$ and $z^{1-\Delta}\Phi$ on the r.h.s. coincide 
with matrix elements of a magnetic translation operator 
realization of the centerless BC superalgebra. (This will be shown in 
Section~\ref{sec4}).

The following bilinear forms come from Eqs.~\eq{H}--\eq{G}:
\beq{Hint}
H_n^{(k)}= -{1\over 2} \oint_0{{dz}\over{2\pi i}}\ :\Psi(z)\, \left(z^{n}
\biggl[ k\Bigl( z\der_{z} +{{n} \over 2}+{1\over2}\Bigr) \biggr]_{-}
\Psi(z) \right): \ ,
\eeq
\beq{Bint}
B_n^{(k)}={1\over 2}\oint_0{{dz}\over{2\pi i}}\,
:\Phi(z)\, \left(z^{n+1}
\biggl[ k\Bigl( z\der_{z} +{n \over 2}+1\Bigr)  \biggr]_{+}
\Phi(z) \right) :\ ,
\eeq
\beqa{Gint}
{G}_r^{(k)}&=& \oint_0{{dz}\over{2\pi i}}\ 
\Psi(z) \, \left( z^{r+{1\over2}}
  q^{-k( z\der_{z} +{r \over 2}+1)}\,\Phi(z) \right) \nn\\
&=& \oint_0{{dz}\over{2\pi i}}\ \Phi(z)\, 
\left( z^{r+{1\over2}} q^{k( z\der_{z} +{r\over 2}+{1\over2})}\,
\Psi(z) \right) \ .
\eeqa
It is convenient to introduce a matrix form for these integral 
representations. If we define the matrix {\Mcal M} as 
\beq{M}
{\mbox{\Mcal M}} ={1\over2}\oint_0{{dz}\over{2\pi i}}\ \chi Q \chi(z)
\eeq
with
\beq{Qz}
\chi(z)=\pmatrix{\Phi(z)&0\cr0&\Psi(z)\cr},\quad
Q(z) =
\pmatrix{z^{n+1}\Bigl[ k\Bigl( z\der_{z} +{n \over 2}+1\Bigr)  \Bigr]_{+}
& z^{r+{1\over2}}q^{k( z\der_{z} +{{r+1} \over 2})}
\cr
z^{r+{1\over2}}q^{-k( z\der_{z} +{{r} \over 2}+1)}
&-z^{n} \Bigl[ k\Bigl( z\der_{z} +{{n+1} \over 2}\Bigr)  \Bigr]_{-}\cr},
\eeq
Eqs.\eq{Hint}--\eq{Gint} are simply described as
\beq{....}
 H_n^{(k)}=\mbox{\Mcal M}_{22}\ ,\quad B_n^{(k)}=\mbox{\Mcal M}_{11}\ ,\quad
{1\over2}\,G_r^{(k)}=\mbox{\Mcal M}_{12}=\mbox{\Mcal M}_{21} \ .
\eeq
By using $\chi$ and $Q$, the commutator representations \eq{Hpsi1}-\eq{Gphi1} can 
be written in some useful integral forms. To show this, we notice that it is 
obvious from Eqs.~\eq{H}--\eq{G} to have the following matrix relation
\beq{matA}
\mbox{\Mcal B}_\chi(w)
\equiv{\pmatrix{[\,B_n^{(k)},\Phi(w)\,]&[\,G_r^{(k)},\Phi(w)\,]\cr
\{\,G_r^{(k)},\Psi(w)\,\}&[\,H_n^{(k)},\Psi(w)\,]\cr}}^T
=\oint{dz\over2\pi i}\pmatrix{z^{n+1}B^{(k)}(z)&z^{r+{1\over2}}G^{(k)}(z)\cr
z^{r+{1\over2}}G^{(k)}(z)&z^{n+1}H^{(k)}(z)\cr}\,\chi(w)\ ,
\eeq
and the r.h.s. is evaluated as the OPE between 
$\chi Q \chi(z)$ and $\chi(w)$. The OPE singular parts of the 
free fields are described as~\cite{OPE2}
\beqa{,1}
&&\Phi(z)\Phi(w) \sim  {1\over{(z-qw)(z-q^{-1}w)}}
={1\over w}\,\bigl[w\der_w\bigr]_-\,{1\over{z-w}}\ ,\\
&&\nn \\
&&\Psi(z)\Psi(w) \sim
{1\over{2}}\left({{q^{1\over2}\over{z-qw}}}
+{{q^{-{1\over2}}}\over{z-q^{-1}w}}\right)
=\bigl[w\der_w+{1\over2}\,\bigr]_+\,{1\over{z-w}}\ ,
\eeqa
or equivalently 
\beq{Pz}
\chi(z)\chi(w)\sim P(z){1\over z-w}\ ;\qquad
P(z)=\pmatrix{-{1\over z}[z\der_z]_-&0\cr0&[z\der_z+{1\over2}]_+\cr}\ .
\eeq
It is also convenient to use the following formulae (the analogues of 
partial integrations) for an analytic function $f(z)$, 
\beqa{partial1,2}
&&\oint{{dz}\over{2\pi i}}\,z^n\,\bigl[k(z\der_z+a)\bigr]_{\pm}f(z)
=\pm\oint{{dz}\over{2\pi i}}\,f(z)\, \bigl[k(z\der_z-a+1)\bigr]_{\pm}
z^n \ ,\label{partial1}\\
&&\oint{{dz}\over{2\pi i}}\,z^n \,q^{k(z\der_z+a)} f(z)
=\oint{{dz}\over{2\pi i}}\, f(z)\,q^{-k(z\der_z-a+1)}
z^n \ .\label{partial2}
\eeqa
Then we obtain \eq{matA} in the following matrix form:
\beq{AQP}
\mbox{\Mcal B}_\chi(w)= \oint_{wq^{\pm1}}{dz\over2\pi i}\,(\chi Q P)(z)\,
{1\over z-w} \ .
\eeq
{}~Further applying \eq{partial1} and \eq{partial2} to each matrix element 
of \eq{AQP}, we also organize it into another different matrix form:
\beqa{APQ.}
\mbox{\Mcal A}_\chi(w)
\equiv{\pmatrix{[\,B_n^{(k)},\Phi(w)\,]&[\,G_r^{(k)},\Phi(w)\,]\cr
-\{\,G_r^{(k)},\Psi(w)\,\}&[\,H_n^{(k)},\Psi(w)\,]\cr}}
&=&- \oint_{w}{dz\over2\pi i}\,{1\over z-w}\, (PQ\chi)(z)\,\nn\\
&=&-PQ\chi(w) \ .\label{APQ}
\eeqa

As shall be shown in Section~\ref{sec5}, the differential operators 
$QP$ and $PQ$ appearing in the representations \eq{AQP} and \eq{APQ} 
are given by the elements of matrix realizations of the centerless 
BC superalgebra in terms of a generalized magnetic translation algebra. 
It is interesting to note that the matrix elements of $P$ disappear 
in the bilinear forms \eq{Hint}--\eq{Gint}, while in \eq{AQP} and 
\eq{APQ}, the matrix $P$ appears as the effect of field contractions. 

\section{The FFZ realizations}\label{sec4}
\setcounter{section}{4}
\setcounter{equation}{0}
\indent


Magnetic translations satisfy the relation (with a suitable normalization)
\beq{ffz1}
 T_{(k,n)}  T_{(l,m)} = q^{ln-mk\over2}(q-q^{-1})^{-1}\, T_{(k+l,n+m)} \ ,
\eeq
and this relation realizes the FFZ algebra~\cite{FFZ}
\beq{ffz2}
 [\,T_{(k,n)}\,,  T_{(l,m)}\,] = \biggl[{ln-mk\over2}\biggr]_- 
T_{(k+l,n+m)} \ .
\eeq
{}~For the moment, we do not specify the forms (realizations) of $T_{(k,n)}$. 
In order to realize a superalgebra we also need the Grassmann operators:
\beq{sigma}
\sigma^2= (\sigma^{\dagger})^2=0\ ,
\quad \{ \sigma, \sigma^{\dagger} \}=1 \ ,
\eeq
which we regard as the quantities commuting with $T_{(k,n)}$, 
\beq{Tsigma}
[T_{(k,n)},\sigma]=[T_{(k,n)},\sigma^{\dagger}]=0\ . 
\eeq
(In Section~\ref{sec5} we discuss the noncommuting case. Cf. Eq.\eq{Tsig}.) 
Using the set of algebras \eq{ffz1}--\eq{Tsigma}, we find the following 
two realizations of the centerless BC superalgebra in accordance with $\pm$ signs 
(we refer to them as ${\cal R}^{\pm}$):
\beqa{RH,B,G}
&&{\hat H}_n^{(k)} = {1\over2}\sum_{\eps,\eta=\pm1}\eta\, 
q^{\pm{\eps n\over2}}T_{(\eta k+\eps,n)}\, \sigma\sigma^\dagger\ ,\\
&&{\hat B}_n^{(k)} = {1\over2}\sum_{\eps,\eta=\pm1}\eps\, 
q^{\pm{\eps n\over2}}T_{(\eta k+\eps,n)}\,\sigma^\dagger\sigma\ ,\\
&&\hat{G}_n^{(k)}=\sqrt{{q-q^{-1}}\over 2}
\sum_{\eps=\pm1}q^{\pm{{\eps n}\over2}} \Bigl(\,\eps^{{1\pm 1}\over2}
T_{(k+\eps,n)}\,\sigma + \eps^{{1\mp 1}\over2}
T_{(-k+\eps,n)}\,\sigma^{\dagger}\,\Bigr)\ .
\eeqa

Now, we specify the $T_{(k,n)}$ operators in terms of the differential 
operators, which possess an additional real parameter $\Delta$; it 
is done by the correspondence 
\beq{Td}
T_{(k,n)}\lra z^nq^{-k(z\der_z+{n\over 2}+\Delta)}/(q-q^{-1})\ .
\eeq
This is not literally the original magnetic translation operator on a 
two-dimensional surface, but actually the dimensionally reduced one~\cite{Hall}. 
It is convenient to rescale the Grassmann operators like 
\beqa{,10}
\sigma = \left(\sqrt{2\over{q-q^{-1}}}\right)^{\pm1}\  
\hat{\sigma}_{1}\ ,\quad 
\sigma^{\dagger} = \left(\sqrt{{q-q^{-1}}\over 2}
\right)^{\mp1}\  \hat{\sigma}_{2} \ ,
\qquad\mbox{for}\quad {\cal R}^\pm\ ,
\eeqa
where $\hat{\sigma}_1$ and $\hat{\sigma}_2$ are still general Grassmann 
operators satisfying Eqs.\eq{sigma} and \eq{Tsigma}. Later we will identify 
$\hat{\sigma}_i$ with Pauli matrices or superspace operators as more 
concrete choices. We then have for the realization ${\cal R}^+$, 
\beqa{,11}
\dand \hat{H}_n^{(k)}=-z^{n}
\left[k\Bigl(z\der_{z} +{n \over 2}+\Delta \Bigr)  \right]_{-}
  \Bigl[z\der_z+\Delta \Bigr]_+\otimes\hat{\sigma}_1\hat{\sigma}_2\ ,\\
\dand \hat{B}_n^{(k)}=-z^{n}
\left[k\Bigl(z\der_{z} +{n \over 2}+\Delta \Bigr)  \right]_{+}
  \Bigl[z\der_z+\Delta \Bigr]_-\otimes\hat{\sigma}_2\hat{\sigma}_1\ ,\\
\dand \hat{G}_n^{(k)}=-z^nq^{-k(z\der_{z}+{n\over 2}+\Delta)}
\Bigl[ z\der_{z}+\Delta \Bigr]_-\hspace*{-6pt}\otimes\hat{\sigma}_1 +
 z^n q^{k(z\der_{z}+{n\over 2}+\Delta)}
\Bigl[ z\der_{z}+\Delta\Bigr]_+\hspace*{-6pt}\otimes\hat{\sigma}_2\ ,
\eeqa
and for the realization ${\cal R}^{-}$,
\beqa{,13}
\dand \hat{H}_n^{(k)}=-z^{n}
\left[k\Bigl(z\der_{z} +{n \over 2}+\Delta \Bigr)  \right]_{-}
         \Bigl[ z\der_{z}+n+\Delta \Bigr]_{+}
\otimes\hat{\sigma}_1\hat{\sigma}_2\ ,\\
\dand \hat{B}_n^{(k)}=-z^{n}
\left[k\Bigl( z\der_{z} +{n \over 2}+\Delta \Bigr)  \right]_{+}
         \Bigl[ z\der_{z}+n+\Delta \Bigr]_{-}
\otimes\hat{\sigma}_2\hat{\sigma}_1\ ,\\
\dand \hat{G}_n^{(k)}=z^nq^{-k(z\der{z}+{n\over 2}+\Delta)}
\Bigl[ z\der_{z}+n+\Delta \Bigr]_+\hspace*{-5pt}\otimes\hat{\sigma}_1-
 z^nq^{k(z\der_{z}+{n\over 2}+\Delta)}
\Bigl[ z\der_{z}+n+\Delta\Bigr]_-\hspace*{-5pt}\otimes\hat{\sigma}_2\ .
\eeqa

Choosing $\hat{\sigma}_i$ as the usual Pauli matrices $\sigma_i$; $i=1,2$,
\beq{Pauli}
\sigma_1=\sigma_x+i\sigma_y=\pmatrix{0&0\cr1&0\cr}\ ,\qquad
\sigma_2=\sigma_x-i\sigma_y=\pmatrix{0&1\cr0&0\cr}\ ,
\eeq
and defining the differential operator matrix in each realization by  
\beq{-2}
\mbox{\Mcal L}^{\pm} ={\hat H}_n^{(k)}+{\hat B}_n^{(k)}+{\hat G}_r^{(k)} \ ,
\qquad\mbox{for}\quad{\cal R}^\pm\ ,
\eeq
we can re-express them as
\beq{**}
\mbox{\Mcal L}^+ = \mbox{\Mcal Q}\mbox{\Mcal P}\ ,
\qquad \mbox{\Mcal L}^- = \mbox{\Mcal P}\mbox{\Mcal Q} \ ,
\eeq
with the matrices similar to $Q$ and $P$ (Cf. Eqs.\eq{Qz} and \eq{Pz})
\beq{*81}
\mbox{\Mcal P}(z)=\pmatrix{-[z\der_z+\Delta]_-&0\cr0&[z\der_z+\Delta]_+\cr}\ ,
\eeq
\beq{*82}
\mbox{\Mcal Q}(z) =
\pmatrix{z^n\Bigl[ k\Bigl( z\der_z +{n \over 2}+\Delta\Bigr)\Bigr]_{+}
& z^r q^{k( z\der_z +{r\over 2}+\Delta)} \cr
z^r q^{-k( z\der_z +{r \over 2}+\Delta)}
&-z^{n} \Bigl[ k\Bigl( z\der_z +{n\over 2}+\Delta\Bigr)\Bigr]_{-}\cr}.
\eeq
Introducing the following notations
\beq{zfield}
\chi'(z)=\pmatrix{\Phi'(z)&0\cr0&\Psi'(z)\cr}\ ,
\qquad \Phi'(z)=z^{1-\Delta}\Phi(z)\ ,\quad
\Psi'(z)=z^{{1\over2}-\Delta}\Psi(z)\ ,
\eeq
we obtain the integral representations similar to \eq{AQP} and \eq{APQ}:
\beq{matB2}
\mbox{\Mcal B}_{\chi'}(w)=\oint{dz\over2\pi i}(\chi'\mbox{\Mcal L}^+)(z)\,
{1\over z-w} \qquad\mbox{(only for $\Delta={1\over2}$)}
\eeq
\beq{matA2}
\mbox{\Mcal A}_{\chi'}(w)= 
- \oint_{w}{dz\over2\pi i}\,{1\over z-w}\, (\mbox{\Mcal L}^- \chi')(z)\,
= - (\mbox{\Mcal L}^- \chi')(w)\ .
\eeq
Apparently, all the matrix elements of \eq{matA2} represent the 
commutation relations \eq{Hpsi2}--\eq{Gphi2}. 

We here put a remark on these two different realizations, which 
are understood as different orderings of {\Mcal P} and {\Mcal Q} 
in the matrix representations. The (differential operator) realization 
${\cal R}^+$ is regarded as a supersymmetric extension of the general 
differential operator expressions presented in~\cite{KS}. However, the 
realization ${\cal R}^-$ is not contained in the literature, and has 
a clear meaning as the adjoint representation \eq{matA2}. Note also 
that \eq{matB2} holds only for $\Delta={1\over2}$, while \eq{matA2} 
holds for arbitrary $\Delta$. 

\section{ The generalized FFZ realizations }\label{sec5}
\setcounter{section}{5}
\setcounter{equation}{0}
\indent 


In order to obtain the differential operator realization 
corresponding to \eq{APQ}, which contains \eq{Hpsi1}--\eq{Gphi1}, 
it may rather be convenient to consider a different magnetic translation 
algebra. {}~For this purpose, let us introduce an additional label 
on $T_{(k,n)}$ correspondingly to the parameter $\Delta$ as introduced 
in \eq{Td}. We then consider the following generalized relations for 
the new sets of 'magnetic translations' 
\beq{gfz1}
 T^{\Delta}_{(k,n)}  T^{\Delta'}_{(l,m)} = 
\frac{q^{{ln-mk\over2}+l(\Delta-\Delta')}}{q-q^{-1}}\ 
T^{\Delta}_{(k+l,n+m)} \ ,
\eeq
and this realizes the generalized FFZ algebra
\beq{gfz2}
 [\,T^{\Delta}_{(k,n)}\,,  T^{\Delta'}_{(l,m)}\,] = 
\frac{q^{{ln-mk\over2}+l(\Delta-\Delta')}}{q-q^{-1}}\ 
T^{\Delta}_{(k+l,n+m)}-
\frac{q^{-{ln-mk\over2}+k(\Delta'-\Delta)}}{q-q^{-1}}\ 
T^{\Delta'}_{(k+l,n+m)} \ . 
\eeq
When $\Delta=\Delta'$, these reduce to the previous relations 
\eq{ffz1} and \eq{ffz2}. We also need the Grassmann operators
\beq{sig}
\tilde{\sigma}^2=(\tilde{\sigma}^{\dagger})^2=0\ ,
\quad \{\tilde{\sigma},\tilde{\sigma}^{\dagger}\}=1,
\eeq
which however do not commute with $T^\Delta_{(k,n)}$:
\beq{Tsig}
T^{\Delta}_{(k,n)}\,\tilde{\sigma}=q^{-{k\over2}}\,\tilde{\sigma}\,
T^{\Delta}_{(k,n)}\ ,\quad
T^{\Delta}_{(k,n)}\,\tilde{\sigma}^{\dagger}=
q^{{k\over2}}\,\tilde{\sigma}^{\dagger}\,
T^{\Delta}_{(k,n)}\ .
\eeq

In this case, the above algebras enable us to have 
different realizations from ${\cal R}^\pm$, 
provided by a certain constraint on $\Delta$ and $\Delta'$; 
hence denoting them as ${\cal R}^\pm_{(\Delta,\Delta')}$. 
We find the following realizations: {}~First, 
we have the realization 
${\cal R}_{(\Delta,\Delta+{1\over2})}^{\pm}$  
with imposing $\Delta'= \Delta + {1\over2}$, 
\beqa{D+H,B,G}
&&\tilde{H}_n^{(k)} = {1\over2}\sum_{\eps,\eta=\pm1}\eta\, 
q^{\pm{\eps n\over2}} T^{\Delta}_{(\eta k+\eps,n)}\,
\tilde{\sigma}\tilde{\sigma}^{\dagger}\ ,\label{D+H} \\
&&\tilde{B}_n^{(k)} = {1\over2}\sum_{\eps,\eta=\pm1}\eps\, 
q^{\pm{\eps n\over2}} T^{\Delta'}_{(\eta k+\eps,n)}\,
\tilde{\sigma}^{\dagger}\tilde{\sigma}\ ,\label{D+B}\\
&&\tilde{G}_n^{(k)}=\sqrt{{q-q^{-1}}\over 2}
\sum_{\eps=\pm1} q^{\pm{{\eps n}\over2}}
\Bigl(\eps^{{1\pm 1}\over2} T^{\Delta}_{(k+\eps,n)}\,
\tilde{\sigma}+\eps^{{1\mp 1}\over2}T^{\Delta'}_{(-k+\eps,n)}\,
\tilde{\sigma}^{\dagger}\,\Bigr)\ ,\label{D+G}
\eeqa
and secondly we have ${\cal R}^{\pm}_{(\Delta,\Delta-{1\over2})}$ 
with imposing $\Delta'=\Delta - 1/2$:
\beqa{D-H,B,G}
&&\tilde{H}_n^{(k)} = {1\over2}\sum_{\eps,\eta=\pm1}\eta\, 
q^{\pm{\eps n\over2}} T^{\Delta}_{(\eta k+\eps,n)}\,
\tilde{\sigma}^{\dagger}\tilde{\sigma}\ , \label{D-H}\\
&&\tilde{B}_n^{(k)} = {1\over2}\sum_{\eps,\eta=\pm1}\eps\, 
q^{\pm{\eps n\over2}} T^{\Delta'}_{(\eta k+\eps,n)}\,
\tilde{\sigma}\tilde{\sigma}^{\dagger}\ ,\label{D-B}\\
&&\tilde{G}_n^{(k)}=\sqrt{{q-q^{-1}}\over 2}
\sum_{\eps=\pm1} q^{\pm{{\eps n}\over2}}
\Bigl(\eps^{{1\pm 1}\over2} T^{\Delta}_{(k+\eps,n)}\,
\tilde{\sigma}^{\dagger}+
\eps^{{1\mp 1}\over2}T^{\Delta'}_{(-k+\eps,n)}\,
\tilde{\sigma}\,\Bigr)\ .\label{D-G}
\eeqa

The differential operator realizations corresponding to these are 
obtained via the replacement
\beq{/3}
T^{\Delta}_{(k,n)}\lra
z^nq^{-k(z\der_z+{n\over 2}+\Delta)}/(q-q^{-1}) \qquad
\mbox{for any $\Delta$}, 
\eeq
if we have the realizations of $\tilde{\sigma}$ and $\tilde{\sigma}^\dagger$, 
which satisfy \eq{sig} and \eq{Tsig} respectively. Note that there are infinitely many 
realizations due to $\Delta$. We hence have to choose suitable realizations case by case, 
in order to make clear connections with other previously given realizations 
and with the $q\ra1$ limits of known forms. In view of this, 
we only discuss the most interesting two cases, which are clearly related 
to the matrices $PQ$ and $QP$ presented in Section~\ref{sec2}.

\subsection{The ${\cal R}^-_{(\Delta,\Delta+{1\over2})}$ case}\label{sec5.1}

In this case ($\Delta'=\Delta+{1\over2}$), we choose the operators 
satisfying \eq{sig} and \eq{Tsig} as 
\beq{choice}
\tilde{\sigma} = z^{1\over2}\sqrt{{q-q^{-1}}\over 2}
\,\hat{\sigma}_{1}\ ,\quad
{\tilde\sigma}^{\dagger}= z^{-{1\over2}}\sqrt{2\over{q-q^{-1}}}
\,\hat{\sigma}_2\ ,
\eeq
where $\hat{\sigma}_i$ are the general operators, which satisfy 
the usual relations \eq{sigma} and \eq{Tsigma}. Then the realizations 
\eq{D+H}--\eq{D+G} are reduced to 
\beqa{/5}
&&\tilde{H}_n^{(k)}= -z^n \left[k\Bigl(z\der_z +{n\over2}+\Delta\Bigr) 
\right]_{-}  \Bigl[ z\der_{z}+n+\Delta \Bigr]_{+}
             \otimes\hat{\sigma}_1\hat{\sigma}_2\ ,\\
&&\tilde{B}_n^{(k)}=-z^n \left[k\Bigl(z\der_z +{n\over2}
          +\Delta+{1\over2}\Bigr) \right]_{+}
 \Bigl[ z\der_{z}+n+\Delta+{1\over2}\Bigr]_{-}
\otimes\hat{\sigma}_2\hat{\sigma}_1\ ,\\
&&\tilde{G}_n^{(k)}=
z^{n+{1\over2}} q^{-k(z\der_{z}+{n\over 2}+\Delta+{1\over2})}
      \Bigl[ z\der_{z}+n+\Delta+{1\over2}\, \Bigr]_+\otimes\hat{\sigma}_1
\nn\\
&&\hspace*{2.5cm}
 -z^{n-{1\over2}} q^{k(z\der_{z}+{n\over2}+\Delta )}
\Bigl[ z\der_{z}+n+\Delta \,\Bigr]_-\otimes\hat{\sigma}_2\ .
\eeqa
If we employ \eq{Pauli} as a realization of $\hat{\sigma}_i$; $i=1,2$, 
with choosing $\Delta={1\over2}$ (and thus $\Delta'=1$), we find 
\beq{.21}
L^-(z)= \tilde{H}_n^{(k)}+\tilde{B}_n^{(k)}+\tilde{G}_r^{(k)}
=PQ(z) \ , \qquad\mbox{for}\quad {\cal R}^-_{({1\over2},1)}.
\eeq
As announced below Eq.\eq{APQ}, each matrix element of $L^-$ provides 
the adjoint representation in \eq{APQ}; i.e., 
\beq{AL}
\mbox{\Mcal A}_\chi(w)=-L^-\chi(w)\ . 
\eeq

We obtain the superspace realization of 
${\cal R}_{(\Delta,\Delta+{1\over2})}^{-}$, 
if we choose the ordinary Grassmann coordinate $\theta$ and its 
derivative $\der_\theta$ as ${\hat\sigma}_i$:
\beq{super1}
  {\hat \sigma}_1=\der_\theta\ ,\qquad  {\hat \sigma}_2=\theta\ .
\eeq
The $q\ra1$ limit should be taken in the combinations (q.v. Eq.\eq{lim}), 
\beq{BHGlim}
H_n^{(k)}/[k]_- +B_n^{(k)} \ra L_n\ , \qquad
G_n^{(k)} \ra G_n\ ,
\eeq
where each limit has a differential realization of the (centerless) 
super-Virasoro algebra: 
\beqa{/8}
&&L_n=-z^n\left(z\der_{z}+{{n+1}\over2}\,
\theta\,\der_{\theta}+{n\over2} +\Delta \right)\ ,\\
&&G_n=z^{n+{1\over 2}}(\der_{\theta}- \theta\der_{z})-
z^{n-{1\over 2}}\theta (n+\Delta) \ .
\eeqa

\subsection{The ${\cal R}^+_{(\Delta,\Delta-{1\over2})}$ case}\label{sec5.2}

In this case ($\Delta'=\Delta-{1\over2}$) we choose 
\beq{/9}
\tilde{\sigma} = z^{{1\over2}}\sqrt{{q-q^{-1}}\over 2}
\,\hat{\sigma}_{2}\ ,\quad
{\tilde\sigma}^{\dagger} = z^{-{1\over2}}\sqrt{2\over{q-q^{-1}}}
\,\hat{\sigma}_1 \ ,
\eeq
and
\beqa{R+Dh,b,g}
&&\tilde{H}_n^{(k)}=-z^{n}
\left[k\Bigl(z\der_{z} +{n \over 2}+\Delta \Bigr)  \right]_{-}
  \Bigl[ z\der_{z}+\Delta \Bigr]_{+}\otimes\hat{\sigma}_1\hat{\sigma}_2\ ,
\label{R+Dh}\\
&&\tilde{B}_n^{(k)}=-z^{n}
\left[k\Bigl( z\der_{z} +{n \over 2}+\Delta-{1\over2} \Bigr)  \right]_{+}
   \Bigl[ z\der_{z}+\Delta-{1\over2} \Bigr]_{-}\otimes\hat{\sigma}_2\hat{\sigma}_1\ ,
\label{R+Db}\\
&&\tilde{G}_n^{(k)}=
-z^{n-{1\over2}}q^{-k(z\der_{z}+{n\over 2}+\Delta-{1\over2})}
\Bigl[ z\der_{z}+\Delta-{1\over2}\, \Bigr]_-\otimes\hat{\sigma}_1\nn\\
&&\hspace*{3.5cm}  +
 z^{n+{1\over2}}q^{k(z\der_{z}+{n\over 2}+\Delta )}
\Bigl[ z\der_{z}+\Delta\,\Bigr]_+\otimes\hat{\sigma}_2\ .
\label{R+Dg}
\eeqa
If we employ \eq{Pauli} as $\hat{\sigma}_i$, with 
$\Delta={1\over2}$ (and thus $\Delta'=0$), we find for 
${\cal R}^+_{({1\over2},0)}$  
\beq{.22}
L^+(z) = \tilde{H}_n^{(k)}+\tilde{B}_n^{(k)}+\tilde{G}_r^{(k)}
=QP(z) \ .
\eeq
As mentioned in the final paragraph of Section~\ref{sec3}, each matrix element 
of $L^+$ provides the adjoint representation in \eq{AQP}; i.e.,
\beq{BL}
\mbox{\Mcal B}_\chi(w)=\int{dz\over2\pi i}(\chi L^+)(z){1\over z-w}\ .
\eeq

The superspace realization  of 
${\cal R}_{(\Delta,\Delta-{1\over2})}^{+}$ can be 
obtained by identifying 
\beq{super2}
  {\hat \sigma}_1=\theta\ ,\qquad  {\hat \sigma}_2=\der_\theta\ ,
\eeq
and the $q\ra1$ limits in \eq{D+H}--\eq{D+G} with \eq{BHGlim} are 
given by 
\beqa{/11}
&&L_n=-z^n\left(z\der_{z}+{{n+1}\over2}\,
\theta\,\der_{\theta}+\Delta-{1\over2}\right)\ ,\\
&&G_n=z^{n+{1\over 2}}(\der_{\theta}- \theta\der_{z})
-z^{n-{1\over2}}\,\theta(\Delta-{1\over2}\,)\ .
\eeqa
In particular, setting $\Delta={1\over2}$ (and $\Delta'=0$), 
we obtain the well-known form
\beqa{./.}
&&L_n=-z^n\left(z\der_{z}+{{n+1}\over2}\,\theta\,\der_{\theta}\right)\ ,\\
&&G_n=z^{n+{1\over 2}}\left(\der_{\theta}- \theta\,\der_{z}\right)\ .
\eeqa

\section{The ordinary field realizations}\label{sec6}
\setcounter{section}{6}
\setcounter{equation}{0}
\indent 

In this section we discuss two sets of the realizations of $H_n^{(k)}$ 
and $B_n^{(k)}$ in terms of ordinary free fields. In the first 
subsection, we present the set (referring to ${H'}_n^{(k)}$ 
and ${B'}_n^{(k)}$) which cannot be related to the deformed field 
realization by changing the normalizations of field oscillators. 
In this sense, this realization may sound to be nontrivial. 
The coefficients in the Sugawara forms only contain rational 
forms of $[x]_\pm$, and we hence call this type the {\it rational} 
realization in this paper. This type takes place in complex (charged) 
free fields. 

On the other hand, the other set (presented in the second subsection) 
is related to the deformed fields by certain normalization changes if 
we examine neutral fields. This type may hence be trivial, {\it however} 
these rescalings of field oscillators give rise to fractional differential 
operations on the ordinary free fields. Also, the coefficients 
in the Sugawara forms are irrational forms of $[x]_\pm$. 
We call this type the {\it irrational} realization.   

\subsection{The rational realizations}\label{sec6.1}
\indent 

As can be seen in \eq{AL} and \eq{BL}, the differential operators $(L^\pm)_{ij}$; 
$i=1,2$, are the adjoint representations in the bases of the deformed 
fields \eq{Psi} and \eq{Phi}. However, as seen in \eq{M}, $(L^\pm)_{ij}$ 
themselves are not exactly the differential operators entered in the 
bilinear realizations \eq{Hint}--\eq{Gint} (Note that the matrix $P$ drops 
out there). Contrastingly, we show that the $P$ matrix part revives in the 
following particular free field bilinear realizations. 

It is well known that by using the ordinary (complex) free fields, 
\beqa{psi,phi}
&&\psi^\ast(z)=\sum_r d^\ast_r\, z^{-r-{1\over 2}}\ ,\quad
\psi(z)=\sum_r d_r\, z^{-r-{1\over 2}},
\qquad \{d_r,d^\ast_s\}=\delta_{r+s,0} \ , \label{psi}\\
&&\der_z\varphi^\ast(z)=\sum_n a^\ast_n\, z^{-n-1}\ ,\quad
\der_z\varphi(z)=\sum_n a_n\, z^{-n-1},
\qquad [a_n,a^\ast_m]=n\,\delta_{n+m,0} \ , \label{phi}
\eeqa
the Virasoro generators are given in the forms 
\beqa{.v}
&&L_n^F= -\int dz\, \psi^\ast(z)\, (z^{n+1}\der_z)\, \psi(z)\ ,\\
&&L_n^B= \int dz\, \der_z\phi^\ast(z)\, (z^{n+1}\der_z)\, \phi(z)\ .
\eeqa
We shall use the abbreviation $\der=\der_z$ as long as it is clear. 
We find that the similar forms
\beqa{.qv}
&&{H'}_n^{(k)}=\int dz\, \psi^\ast(z)\, (L^+)_{22}\, \psi(z)\ ,\\
&&{B'}_n^{(k)}=-\int dz\, \der\phi^\ast(z)\, (L^+)_{11}\, \phi(z)
\eeqa
satisfy the same algebra as \eq{BCalg} with having a factor of 2 in the 
central extensions given in the cases of $H_n^{(k)}$ and $B_n^{(k)}$ 
(see \eq{CB} and \eq{CH}). Obviously in these realizations the matrix 
elements of $P$ participate in 
\beqa{Hnk1,2}
{H'}_n^{(k)} &=& -\oint{dz\over2\pi i}\, \psi^\ast(z) z^n
[z\der+{1\over2}]_+\,[k(z\der+{1\over2}(n+1))]_-\,\psi(z) \label{Hnk1}\\
&=& - \sum_{r\in {\z}+1/2} [k({n\over2}-r)]_-
[r]_+:d^\ast_{n-r}d_r:\ , \label{Hnk2}
\eeqa
and
\beqa{Bnk1,2}
{B'}_n^{(k)} &=& \oint{dz\over2\pi i}\, \der\varphi^\ast(z)z^n
[k(z\der+{n\over2})]_+ [z\der]_-\,
\varphi(z) \label{Bnk1} \\
&=& \sum_{l\in {\z}} {[l]_-\over l}
[k({n\over2}-l)]_+:a^\ast_{n-l}a_l:\ . \label{Bnk2}
\eeqa
Applying the formula \eq{partial1} to \eq{Hnk1} and \eq{Bnk1}, 
we also derive another forms associated to $(L^-)_{ii}$: 
\beq{Hint0}
{H'}_n^{(k)}= - \oint{dz\over2\pi i}\, L^-_{22} \,
\psi^\ast(z)\cdot\psi(z)\ ,
\eeq\beq{Bint0}
{B'}_n^{(k)}= - \oint{dz\over2\pi i}\, L^-_{11} \,
\der\phi^\ast(z)\cdot\phi(z)\ .
\eeq
Here, the $L^-$ can thus be understood as a partially integrated 
version of $L^+$. 
As we can see in \eq{Hnk2} and \eq{Bnk2}, ${H'}_n^{(k)}$ and 
${B'}_n^{(k)}$ are not related to \eq{H} and \eq{B} by changing any 
normalizations of the oscillators. (Although we are using complex fields here, 
it is clear that the argument is straightforward.) Therefore, in this case, 
the deformed fields \eq{Psi} and \eq{Phi} cannot be interpreted in terms of a 
normalization change from the ordinary fields \eq{psi} and \eq{phi}.

As a connection to the next section (irrational realizations), let us 
consider a fractional power decomposition of a $q$-derivative. 
{}~For an analytic function $f(x)$, defining 
\beq{def1}
f(z\der) z^n \define f(n) z^n \ ,
\eeq
we observe
\beqa{irPsi}
&&\sqrt{[z\der_z+{1\over2}]_+} \sqrt{[w\der_w+{1\over2}]_+}
<\psi^\ast(z)\psi(w)>=
[z\der_z+{1\over2}]_+<\psi^\ast(z)\psi(w)>\nn\\
&&=[w\der_w+{1\over2}]_+<\psi^\ast(z)\psi(w)> \ ,
\eeqa
and
\beqa{irPhi}
&&\sqrt{[z\der_z+1]_-\over z\der_z+1}\sqrt{[w\der_w+1]_-\over w\der_w+1}
<\der\varphi^\ast(z)\der\varphi(w)>
= {[z\der_z+1]_-\over z\der_z+1}<\der\varphi^\ast(z)\der\varphi(w)>\nn\\
&&= {[w\der_w+1]_-\over w\der_w+1}<\der\varphi^\ast(z)\der\varphi(w)>
={1\over w}[w\der_w]_-<\der\varphi^\ast(z)\varphi(w)>\ .
\eeqa
Thus, roughly speaking, the pair of $\psi^\ast(z)$ and 
$[z\der+{1\over2}]_+\psi(z)$ in \eq{Hnk1}, and the pair of 
$\der\phi^\ast(z)$ and $z^{-1}[z\der]_-\phi(z)$ in \eq{Bnk1} may be 
replaced by bilinear forms of the following new fields:
\beq{ifield}
{\tilde \Psi}(z)=\sqrt{[z\der_z+{1\over2}]_+}\,\psi(z)\ ,\quad
{\tilde \Phi}(z)=\sqrt{[z\der_z+1]_-\over z\der_z+1}\,\der\varphi(z)\ ,
\eeq
and similarly for ${\tilde \Psi^\ast}(z)$ and ${\tilde \Phi^\ast}(z)$. 

As shall be seen in the next subsection, this construction serves 
different realizations, however satisfying the same algebra 
(without changing the central extensions). These fractional 
nonlocal operations \eq{ifield} lead to simple normalization 
changes in the Fourier mode oscillators $d_r$ and $a_n$. 

\subsection{The irrational realizations}\label{sec6.2}
\indent 

In this section we consider normalized forms inferred from the previous section, 
particulaly through the argument from Eq.\eq{irPsi} to Eq.\eq{ifield}. 
It is convenient to return to neutral field cases in respect to comparison 
with Section~\ref{sec3}. It is very natural to expect that the 
relations \eq{ifield} would also apply to the neutral fields, 
and we start from 
\beqa{Hnk3,B}
&&H_n^{(k)} = -{1\over2}\oint{dz\over2\pi i}\,\tilde{\Psi}(z)\,z^n
\,[k(z\der+{1\over2}(n+1))]_-\,\tilde{\Psi}(z)\ , \label{Hnk3}\\
&&B_n^{(k)} = {1\over2} \oint{dz\over2\pi i}\,\tilde{\Phi}(z)\,
z^{n+1}\,[k(z\der+{n\over2}+1)]_+\,\tilde{\Phi}(z)\ . \label{Bnk3} 
\eeqa
Easily understood from definitions \eq{def1} and \eq{ifield}, the Sugawara 
forms of these contain the irrational (root) forms of $[x]_\pm$; i.e.
\beq{irH}
H_n^{(k)}={1\over 2}\sum_r\Bigl[k({n\over 2}-r)\Bigr]_-
\sqrt{{[r]_+[n-r]_+}}\ :d_r d_{n-r}: \ ,
\eeq
\beq{irB}
B_n^{(k)}={1\over 2}\sum_j\Bigl[k({n\over 2}-j)\Bigr]_+
\sqrt{{[j]_-[n-j]_-}\over{j(n-j)}}\ :a_j a_{n-j}:\ .
\eeq
Obviously, these cannot be transformed into the rational 
realizations \eq{Hnk2} and \eq{Bnk2} by performing 
$d_r \ra f(n,r)d_r$ etc. with any function $f$.

It is worth noticing that the bilinear field forms for these 
realizations are certainly the same as the deformed ones~\eq{Hz} 
and \eq{Bz}:
\beq{irHz}
H^{(k)}(z)=\sum_n H_n^{(k)}z^{-n-2} 
={1\over z(q-q^{-1})}{\tilde\Psi}(q^{k/2}z)
{\tilde\Psi}(q^{-k/2}z)\ , 
\eeq
\beq{irBz}
B^{(k)}(z)=\sum_n B_n^{(k)}z^{-n-2}
={1\over 2}{\tilde\Phi}(q^{k/2}z){\tilde\Phi}(q^{-k/2}z)\ .
\eeq
In fact, Eqs.\eq{irH} and \eq{irB} coincide with the 
rescaled objects obtained from \eq{H} and \eq{B}, if 
identified by the relations 
\beq{scale}
\sqrt{[r]_+}\,\,d_r=\Frb_r\ ,\qquad
\sqrt{[n]_-\over n}\,\,a_n = \Fra_n\ ,
\eeq
where $\Frb_r$ and $\Fra_n$ are the deformed oscillators defined 
in \eq{Psi} and \eq{Phi}. This identification also matches with 
the supercurrent generators $G^{(k)}(z)$, which are given by 
\beq{..10}
G^{(k)}(z) = q^{-{k\over 4}}{\tilde\Psi}(q^{k/2}z)
{\tilde\Phi}(q^{-k/2}z) =\sum_r G_r^{(k)}z^{-r-{3\over 2}}\ ,
\eeq
where
\beq{.0.}
G_r^{(k)}=\sum_j q^{-k({r\over 2}-j)}
\sqrt{{[j]_-[r-j]_+}\over{j}}\ a_j d_{r-j}\ .
\eeq
The set of $H_n^{(k)}$, $B_n^{(k)}$ and $G_r^{(k)}$ 
given in this realization of course satisfies the 
same superalgebra as given in Section~\ref{sec2}. This is the 
reason why we have used the same notation as used in 
Section~\ref{sec3}.

\section{The $bc$ field representation}\label{sec7}
\setcounter{section}{7}
\setcounter{equation}{0}
\indent

In this section, to find some more realizations for the algebra \eq{BCalg}, 
we assume the general commutation relation (including the previous cases 
of deformed/undeformed bosons and fermions):
\beq{.2.}
 c_n b_m +\epsilon b_m c_n = D_n\,\delta_{n+m,0} \ ,
\eeq
where $b_n$ and $c_n$ are interpreted as the usual $bc$ ghosts ($h=2$) 
for $\eps=1$, and as the $\beta\gamma$ superghosts ($h=3/2$) for 
$\eps=-1$. The previous bosons ($h=0$) and fermions ($h={1\over2}$) are 
identified to the cases that $c_n=d_n$, $a_n$ and $b_n=d^\ast_n$, $a^\ast_n$. 
The $h$ stands for the conformal dimension of the field 
corresponding to $b_n$ (at $q=1$) in each case, and the normalization 
$D_n$ should be given case by case. 

Let us consider the following bilinear form with the coefficients 
$s_n^l(k)$ yet to be determined from the consistency of \eq{BCalg}: 
\beq{.1.}
E_n^{(k)} = \sum_{l} s_n^l(k) :b_{n-l}c_l:\ , \qquad (l\in {\Z}-h). 
\eeq
We calculate the commutation relation for $E_n^{(k)}$ 
\beqa{formula1}
&&[\,E_n^{(i)},\, E_m^{(j)}\,] = 
\sum_{l}\Bigl( s_m^l(j) s_n^{l-m}(i) D_{l-m}-
s_m^{l-n}(j) s_n^l(i) D_{l-n}\Bigr):b_{n+m-l}c_l: \nn\\
&&+\epsilon\delta_{n+m,0}\sum_{0<l\leq n}
s_{-n}^{-l}(j) s_n^{n-l}(i)D_{n-l}D_{-l}\ .
\eeqa

Examining this formula for the deformed $bc$ and $\beta\gamma$ cases 
with $D_n =[n]_+$, we find that the following form 
\beq{Cnk}
E_n^{(k)} = -\sum_{l\in {\z}-h} [k({n\over2}-l)]_-
:b_{n-l}c_l:\ 
\eeq
satisfies the algebra \eq{BCalg} with the central extensions given by 
\beq{.4.}
C_E= \eps\sum_{0<l\leq n} [j({n\over2}-l)]_- [i({n\over2}-l)]_-
[l-n]_+ [l]_+ \ .
\eeq
On the analogy of \eq{Phi}, instead of \eq{Cnk}, it is also obvious 
to have different realizations with choosing $D_n=[n]_-$; 
\beq{C'nk}
{E'}_n^{(k)} = \sum_{l\in {\z}-h} [k({n\over2}-l)]_+
:b_{n-l}c_l: \ ,
\eeq
however the values of central extensions differ from the above 
case
\beq{.6.}
C_{E'}= \eps\sum_{0<l\leq n} [j({n\over2}-l)]_+ [i({n\over2}-l)]_+
[l-n]_- [l]_- \ .
\eeq

One may further obtain some ordinary field realizations (without 
making any changes in the central extensions up to the factor of 2), 
examining the formula \eq{formula1} with setting $D_n=1$. 
{}~For example, we find a third boson realization for $B_n^{(k)}$, 
\beq{newB} 
{B''}_n^{(k)} = \sum_{l\in{\z}}[k({n\over2}-l)]_+[l]_-
:\alpha^\ast_{n-l}\alpha_l:
\eeq
with 
\beq{.8.}
[\alpha_n,\alpha^\ast_m]=\delta_{n+m,0} \ ,
\eeq
which is related to Eq.\eq{phi} by the normalization relation
\beq{.9.}
a_n = \sqrt{|n|}\,\, \alpha_n \ ,\qquad 
a^\ast_n = \sqrt{|n|}\,\, \alpha^\ast_n \ .
\eeq
One can see that this realization, 
\beq{.10.} 
{B''}_n^{(k)} = \sum_{l\in{\z}}[k({n\over2}-l)]_+
{[l]_-\over \sqrt{|l(n-l)|}}:a^\ast_{n-l}a_l:\ ,
\eeq
differs from the previous realizations \eq{Bnk2} and \eq{irB}. 

In the similar way to \eq{newB}, we find the realizations 
\beq{Cnk2}
E_n^{(k)} = -\sum_{l\in {\z}-h} [k({n\over2}-l)]_-[l]_+
:b_{n-l}c_l:\ , \qquad \mbox{with} \quad D_n=1
\eeq
\beq{C'nk2}
{E'}_n^{(k)} = \sum_{l\in {\z}-h} [k({n\over2}-l)]_+[l]_-
:b_{n-l}c_l: \ , \qquad \mbox{with} \quad D_n=1.
\eeq 
However, these are equivalent to \eq{Cnk} and \eq{C'nk} after 
rescaling not $b_n$ but $c_n\ra [n]_\pm c_n$ respectively. This is 
a different feature from the boson and fermion cases.

\section{Conclusions and discussions}\label{sec8}
\setcounter{section}{8}
\setcounter{equation}{0}
\indent

We have studied various realizations for the superalgebra proposed by 
Belov and Chaltikian~\cite{BC} as a deformation of the Virasoro algebra 
in terms of: Moyal-like operators, differential operators, the adjoint 
commutators and their matrix representations, the bilinear integral forms 
of deformed/nondeformed fields. In Sections~\ref{sec4} and \ref{sec5}, 
we first have constructed the deformed (centerless) Virasoro operators 
at the abstract level based on the noncommuting Moyal type operators, 
and then have transformed them into differential operators realizations 
(in other words, $q$-differences or reduced magnetic translations). 
The superspace realizations are straightforward. All these are an infinite 
number of realizations in accordance with a parameter $\Delta$. At these 
differential operator levels, we do not have to specify the value of $\Delta$. 

While considering a connection of them to realizations in field theory, 
it has been necessary to specify $\Delta$ (Sections~\ref{sec5} and \ref{sec6}). 
The adjoint matrix forms derived in Section~\ref{sec3} and the differential 
operator forms $L^\pm$ in Section~\ref{sec5} have played a key role to find 
this connection, where the same $PQ$ matrix combinations appear. 
In this way, all the formulae presented in Section~\ref{sec3} are 
reproduced for special values of $\Delta$. The similar statement 
holds for the results of Section~\ref{sec4}, however, it is interesting 
to note that \eq{matA2} holds without specifying the $\Delta$ value. 

Let us compare in details the results between Sections~\ref{sec4} 
and \ref{sec5}. In Section~\ref{sec4}, we treated 
the Grassmann variables as quantities commuting with the FFZ parts, 
while in Section~\ref{sec5} we treated them as those noncommuting with the FFZ 
parts. This noncommutativity is realized by the Pauli matrices multiplied 
by $z^{\pm1/2}$ in the differential realizations (see \eq{choice}), while 
these multipliers are absorbed in the field definitions \eq{zfield} in 
the former case. This is the reason why we encounter the unusual versions 
of adjoint commutator representations \eq{Hpsi2}--\eq{Gphi2} 
(Cf. Eqs. \eq{Hpsi1}--\eq{Gphi1}). Due to this fact, the $PQ$ and 
{\Mcal P}{\Mcal Q} matrix formulations become slightly different 
from each other (qv. \eq{AQP}, \eq{APQ}, \eq{matB2} and \eq{matA2}). 
These two sets of realizations are very similar at this stage, 
however it is contrast that their original abstract constructions are 
very different. As mentioned above, the arbitrariness of 
$\Delta$ in \eq{matA2} is a big difference as well. 

Although we did not mention explicitly that the bilinear form \eq{M} can be 
reproduced from our differential operators, it is also possible to do. 
Once we read the matrix $Q(z)$ from $L^\pm(z)$ by removing $P(z)$, 
we have only to sandwich $Q$ by a couple of deformed/undeformed fields. 
This structure can be observed from \eq{Hnk3} and \eq{Bnk3} in the case 
of undeformed field case. The deformed field case is straightforward. 
The similar statement holds for the results of Section~\ref{sec4} with 
taking account of the $z$ multiplications into the $\chi'$ fields. 
After all, the matrix factor $P$ plays a role of deforming propagators 
between two fields. 

In Section~\ref{sec6}, we have found the realizations, making use of the 
usual (nondeformed) free fields. The relations of the irrational 
realizations to the deformed field realizations are simply normalizations 
at the level of oscillators, however they are nontrivial at the level of 
fields. This fact suggests that a noncommutative string could be described 
in terms of either simple deformed fields, or highly nontrivial operations 
on ordinary fields. Judging from these observations, the relations among 
the deformed field and their corresponding {}~FFZ and differential 
realizations are the most significant of all the results we found in this 
paper. There then comes a question whether a physical system satisfying 
the relations \eq{gfz1}--\eq{Tsig} exists or not.

Concerning Section~\ref{sec7}, we obtained the same copies of the BC 
superalgebra in terms of (deformed/nondeformed) ghost oscillators with 
new central extensions. In a $n$ dimensional system like a string theory, 
the cancellation of the central extensions $n(C_B+C_H)+C_E=0$ is possible 
for a certain value of $q$. In the language of string theory, $n$ means 
the critical dimensions, where anomalies vanish. Unfortunately the present 
algebra have a problem to apply this argument; i.e., the $q\ra1$ limit is 
different from the usual Virasoro operators for the ghost parts. However, 
the idea of vanishing anomaly at a special value of $q$ has become more 
promising as a model of deformed (noncommutative) superstring than ever. 
Similar applications of the central charges would be possible in any other 
models possessing the Virasoro algebra. 

It is clear that our deformation is related to the Moyal 
quantization since we started from the FFZ realizations. In this sense 
also, the present deformed super Virasoro algebra should thus be understood 
as a noncommutative deformation of the original Virasoro algebra. We believe 
that it is important to further pursue the relations between our results 
and recent developing noncommutative physics and field theories. 
The representation theory should also be studied along the 
similar line. 

\par
{\bf Acknowledgements}
The authors would like to thank the members of Theory Group, KEK Tanashi 
for their kind hospitality. H.T.S. has been supported by the Brain Pool 
Program (KOSEF) and by the BK 21 project. 



\begin{thebibliography}{99}

\bibitem{saito} S.Saito, Integrability of Strings, in: Nonlinear Fields: 
Classical, Random, Semiclassical, eds. P.Garbaczewski and Z.Popowicz \WS{1991} 
p.286; $q$-Virasoro and $q$-Strings, in: Quarks, Symmetries and Strings, 
eds. M.Kaku, A.Jevicki and K.Kikkawa \WS{1991} p.231;\\
H. Hiro-oka, O. Matsui, T. Naito and S. Saito, preprint 
TMUP-HEL 9004(1990), unpublished.
\bibitem{CP1} M. Chaichian and P.P. Pre\v{s}najder, \PL{B277} (1992) 109.
\bibitem{KS}R. Kemmoku and S. Saito, \PL{B319} (1993) 471.
\bibitem{BC} A.A. Belov and K.D. Chaltikian, \MPL{A8} (1993) 1233.
\bibitem{N2} E. Batista, J.F. Gomes and I.J. Lautenschleguer, 
\JP{A29} (1996) 6281. 

\bibitem{QST} M.Chaichian, A.Demichev and P.Pre\v{s}najder, preprint 
HIP-2000-08/TH (hep-th/0003270).
\bibitem{QKZ}S. Lukianov and Y. Pugai, 
{\it J. Exp. Theor. Phys.} {\bf 82} (1996) 1021; 
S. Lukianov, \PL{B367} (1996) 121.
\bibitem{me} A. Jellal and H-T. Sato, \PL{B483} (2000) 451.
\bibitem{Hall} H-T. Sato, \ZP{C70} (1996) 349; 
\PTP{93} (1995) 195 (hep-th/9312174). 
\bibitem{KP} R. Kemmoku and S. Saito, \JP{A29} (1996) 4141; 
(see also hep-th/9411027).

\bibitem{CH} C.S. Chu and P.M. Ho, \NP{B550} (1999) 151.
\bibitem{SW} N.Seiberg and E.Witten, JHEP 9909 (1999) 32.

\bibitem{mag} S. M. Girvin, A. H. MacDonald and P. M. Platzman, 
      \PR{B33} (1986) 2481.
\bibitem{FFZ} D.B. Fairlie, P. Fletcher and C.K. Zachos, 
\PL{B218} (1989) 203; \JMP{31} (1990) 1088; 
D.B. Fairlie and C.K. Zachos, \PL{B224} (1989) 101.
\bibitem{moyal} F. Bayen, M. Flato, C. Fronsdal, A. Lichnerowicz 
and D. Sternheimer, \AP{111} (1978) 61; ibid 111.
\bibitem{NC} D.B. Fairlie, \MPL{A13} (1998) 263;\\
E.G. Floratos and G.K. Leontaris, \PL{B464} (1999) 30; \\
I. Bars and D. Minic, preprint USC-99/HEP-B5 (hep-th/9910091);\\
J. Madore, S. Schraml, P. Schupp and J. Wess, LMU-TPW-2000-05 (hep-th/0001203).

\bibitem{OPE1} H. Sato, \NP{B393} (1993) 442.
\bibitem{OPE2} H-T. Sato, \NP{B471} (1996) 553.
\bibitem{CP2} M. Chaichian and P. Pre\v{s}najder, 
             \NP{B482} (1996) 466 (see also hep-th/9603064). 

\bibitem{LLL} I.I. Kogan, \IJMP{A9} (1994) 3887;\\
H-T. Sato, \MPL{A9} (1994) 451; {\it ibid.} 1819; \MPL{A10} (1995) 853.

\end{thebibliography}
\end{document}